\documentclass[preprint,preprintnumbers,amsmath,amssymb]{revtex4-1}

\usepackage{graphicx,tabularx}% Include figure files
\usepackage{bm}
\usepackage{float}
\usepackage{dcolumn}% Align table columns on decimal point
\usepackage{color}
\usepackage{ulem}
\usepackage{moreverb}
\usepackage{amsmath}
\usepackage{setspace}

\renewcommand{\thefigure}{\arabic{figure}}


\makeatletter
\def\blfootnote{\xdef\@thefnmark{}\@footnotetext}
\makeatother

\begin{document}
\renewcommand{\figurename}{\textbf{Fig.}}
\renewcommand{\thefigure}{\textbf{\arabic{figure}}}

\title{Cascade of magnetic-field-induced quantum spin states in a spin-1 honeycomb magnet}

\author{Kaixin Tang$^{1,\star}$}
\author{Zhao-Yang Dong$^{2,\star}$}
\author{Yanyan Shangguan$^{2,3,\star}$}
\author{Zhao Gong$^{1}$}
\author{Ye Yang$^{4}$}
\author{Song Bao$^{3}$}
\author{Houpu Li$^{5}$}
\author{Nan Zhang$^{5}$}
\author{Hongyu Li$^{1}$}
\author{Jian-Xin Li$^{3,6}$}
\email{jxli@nju.edu.cn}
\author{Jinsheng Wen$^{3,6}$}
\email{jwen@nju.edu.cn}
\author{Ziji Xiang$^{1,7}$}
\email{zijixiang@ustc.edu.cn}
\author{Xianhui Chen$^{1,5,7}$}
\email{chenxh@ustc.edu.cn}

\affiliation{
	\vspace{10pt}
	$^1$ Hefei National Research Center for Physical Sciences at the Microscale, University of Science and Technology of China, Hefei, Anhui 230026, China\\
	$^2$ School of Physics, Nanjing University of Science and Technology, Nanjing 210094, China\\
	$^3$ National Laboratory of Solid State Microstructures and School of Physics, Nanjing University, Nanjing 210093, China\\
	$^4$ School of Physics, Hefei University of Technology, Hefei, Anhui 230009, China\\
	$^5$ Department of Physics, University of Science and Technology of China, Hefei, Anhui 230026, China\\
	$^6$ Collaborative Innovation Center of Advanced Microstructures and Jiangsu Physical Science Research Center, Nanjing University, Nanjing 210093, China\\
	$^7$ Hefei National Laboratory, University of Science and Technology of China, Hefei, Anhui 230088, China.
}
\date{\today}
%\pacs{71.18.+y, 74.25.Ha, 74.25.Jb}
\blfootnote{$\star$These authors contributed equally to this work.}

\begin{singlespacing}
	\maketitle                 % Produces the title
\end{singlespacing}

%Abstract
\noindent
\textbf{Quantum fluctuations endow spin systems with surprisingly enriched magnetic phase diagrams. In frustrated magnets, strong quantum fluctuations boosted by either geometrical incompatibility or competitive exchange interactions stabilize cascades of unusual phases of matter. Here we reveal the presence of multiple quantum phases in the honeycomb antiferromagnet Na$_{3}$Ni$_{2}$BiO$_{6}$, both inside and beyond its field-induced one-third magnetization plateau. Comprehensive measurements of thermodynamic quantities demonstrate that the one-third plateau comprises at least three distinct spin states with nearly-degenerate net magnetization, separated by first-order transitions that likely involve sequential spin reconfiguration. Upon further increasing the magnetic field, the system evolves across a myriad of peculiar phases before reaching full polarization; these intermediate phases possess copious low-energy excitations, manifested by anomalous upturns of specific heat at ultralow temperatures — probably hinting at the development of ``hidden" ordered ground states. The complex magnetic phase diagram of Na$_{3}$Ni$_{2}$BiO$_{6}$ underlines the preponderant impact of quantum fluctuations on a honeycomb spin lattice with strong exchange frustration.
}

%Introduction
\bigskip
\noindent
\textbf{\large Introduction}

\noindent
Frustrated magnets offer a blue-print to realize fascinating quantum states and emergent phenomena. The frustrations arising from magnetic lattice geometry (i.e., the geometric frustration) \cite{RVB_Kalmeyer,OngCavaTriangular} or from competing magnetic interactions (i.e., the exchange or bond frustration) \cite{SquareFrus}
prevail in different types of spin systems, yet both explicitly enhance the quantum fluctuation in the spin degree of freedom \cite{StarykhReview}. Whilst abundance of quantum fluctuations potentially leads to the spin liquid state that remains disordered down to zero temperature ($T$) \cite{BalentsSL,YueshengSL}, in most cases the spin systems establish long-range orders at low $T$ and, with the assistance of quantum fluctuations, develops a plethora of unconventional quantum phases such as fractional magnetization plateaus \cite{ChubukovPlateau,Zhitomirsky,Cs2CuBr4Plateau,Cs2CuBr4Forture,Ba3CoSb2O9_MH}, longitudinal spin-density waves \cite{LongitudinalSDW}, supersolid phase \cite{Supersolid,MnCr2S4,NaBaCoPO_MCE}, spin-nematic state \cite{SpinNematic,NaBaNiPO}, etc, under the application of magnetic fields ($H$) --- some of those are classically unexpected \cite{StarykhReview,Cs2CuBr4Forture}. Among the frustrated magnets, the quasi-two-dimensional (2D) triangular lattice antiferromagnets are of special interest because their geometric frustration promises (i) strong quantum fluctuations, which stabilize a unique up-up-down ground state manifested by a magnetization plateau at one-third of the saturation value (the 1/3-plateau phase hereafter) in a certain range of $H$ \cite{StarykhReview,ChubukovPlateau,Coletta2013,XXZmodel}; (ii) unusual magnon condensates that are analogue to the supersolid state emerging from hard-core bosons on a triangular lattice \cite{Supersolid,XXZmodel,Supersolid2}. In principle, the density of bosonic spin excitations in the magnetically ordered states (i.e., magnons) can be effectively manipulated by the external $H$-field through the Zeeman-type coupling \cite{BEC}; subsequently, the $H$-field drives the system through a quantum critical point (QCP) marking an analogous of the Bose-Einstein condensation (BEC) of magnons \cite{BEC_review} and triggers intriguing critical behaviour in its vicinity \cite{NaBaCoPO,YbCl3}. Latest experimental results have revealed that in quasi-2D triangular lattice antiferromagnets Na$_2$BaCo(PO$_4$)$_2$ and Na$_2$BaNi(PO$_4$)$_2$, a BEC-QCP occurs beyond the endpoint of the 1/3-plateau phase in the magnetic phase diagram; in the former, the BEC leads to spin-lattice supersolidity characterized by remarkable low-$T$ entropy stemming from gapless Goldstone modes (corresponding to the the U(1) phase fluctuations) \cite{NaBaCoPO_MCE}, whereas the latter, successfully described by a spin-1 model, develops a two-magnon BEC that gives rise to a spin-nematic state \cite{NaBaNiPO}.

In quasi-2D honeycomb spin lattices, the above-mentioned exotic magnetically ordered phases are not inherent, because the spin frustration can hardly be established without involving complex intersite exchange interactions. Nevertheless, a spin-1 honeycomb lattice antiferromagnet Na$_3$Ni$_2$BiO$_6$ has recently been identified as a notable exception \cite{Na3Ni2BiO6_plateau}. This compound crystallizes in a layered monoclinic structure (space group $C$2/m) comprising stacks of alternating Ni-Bi-O honeycomb layers and Na atom layers along the perpendicular $c^*$ axis; the Ni atoms adopt a honeycomb geometry in the $ab$ plane (See Supplementary Fig.\,1a) \cite{Na3Ni2BiO6_plateau,Na3Ni2BiO6_structure}. The antiferromagnetic (AFM) order sets in at the N\'{e}el temperature $T_{\rm N} \approx$ 10 \,K \cite{Na3Ni2BiO6_structure}, characterized by an in-plane zigzag configuration of the Ni$^{2+}$ spins ($S$ = 1), which are aligned along the monoclinic $c$ axis ($\sim$ 19$^\circ$ away from the perpendicular direction $c^*$) \cite{Na3Ni2BiO6_plateau,Na3Ni2BiO6_NMR}. With the application of $H$ along $c^*$, a 1/3-magnetization plateau emerges at low $T$ between $\mu_0H \approx$ 5.2 and 7.4\,T, and a spin-polarized state is realized at higher $H$ (8.3\,T at 2\,K) \cite{Na3Ni2BiO6_plateau}. It has been proposed that the bond-anisotropic Kitaev interaction is an essential prerequisite for the occurrence of 1/3-magnetization plateau \cite{Na3Ni2BiO6_plateau}: the inclusion of Kitaev interaction creates pronounced exchange frustrations and, consequently, strong quantum fluctuations. The underlying mechanism of the 1/3-plateau phase in Na$_3$Ni$_2$BiO$_6$ can be different from the order-by-disorder scenario (that quantum fluctuations select one of the classically-degenerate spin configurations in a finite field range) \cite{Na3Ni2BiO6_plateau} that is frequently referred to the cases of geometrically frustrated magnets, e.g., triangular or kagome lattices \cite{StarykhReview,ChubukovPlateau,Zhitomirsky,OrderByDisorder}. However, the nature of this intriguing phase remains elusive.

Here, we investigate the field-induced quantum phase transitions and spin states in Na$_3$Ni$_2$BiO$_6$ from the thermodynamic perspective using four complementary experimental probes: ac magnetostriction, dc magnetization, torque magnetometry and specific heat. We track the evolution of thermodynamic properties under magnetic fields up to 14\,T down to an ultralow temperature of 0.1\,K and map the magnetic phase diagram, which appears to be surprisingly complicated. The 1/3-magnetization plateau is determined to be composed of three distinct phases down to $\sim$ 0.35\,K and probably more at even lower $T$, instead of being a single phase; such phenomenon has seldom been observed in geometrically frustrated spin lattices, suggesting the highly unusual nature of the fractional plateau states herein. Above the critical $H$-field marking the endpoint of 1/3 magnetization plateau, successive quantum phase transitions occur before the realization of full magnetization saturation. The corresponding quantum spin phases, termed intermediate phases in this work, exhibit abnormal increases of specific heat $C_{\rm p}(T)$ upon cooling at the lowest $T$: $C_{\rm p}(T)/T \propto T^{-\alpha} (\alpha \simeq 2.4-3.0)$, which cannot be attributed to the conventional Schottky anomalies; this inverse-power-law rise contradicts the theoretical expectations for gapless magnetic excitations and most likely implies the incipient signature of a transition/crossover towards a ``hidden" ground state (residing below 100\,mK) stemming from the in-plane component of Ni spins. The overall phase diagram bears some degree of resemblance to certain quasi-2D triangular lattice antiferromagnets \cite{NaBaCoPO_MCE,NaBaNiPO}, yet the quantum spin states can be completely different from their counterparts in essence. The segmented 1/3-plateau phases and the intermediate phases with masked ground states both emerge uniquely from the strong quantum fluctuations prompted by exchange frustration. These results highlight the fundamental role of Kitaev physics in promoting unusual quantum phenomena in honeycomb magnets.

%Experimental Results
\bigskip
\noindent
\textbf{\large Results}

\noindent
\textbf{Magnetic-field-induced quantum phase transitions}

\noindent
We explored the field-induced phase transitions in Na$_{3}$Ni$_{2}$BiO$_{6}$ by the method of ac { magnetostriction} measurement \cite{MES_MnSi,MES_2Dmagnets,MES_QOs}. This technique directly probes the linear magnetostriction coefficient $\lambda(H) = (1/L_0)\partial\Delta L(H)/\partial\mu_0H$ (here $L_0$ is the sample length at zero $H$ and $\Delta L$ is the variation of sample length under $H$) by converting it to an ac electrical voltage signal in a magnetoelectric (ME) composite configuration, as illustrated in Fig.\,1a. More information of the measurement is presented in Methods. Figure 1b displays the $H$-dependent real part of the ME voltage coefficient, $\alpha_{x}(H)$, a parameter that is proportional to $\lambda(H)$ \cite{MES_MnSi,MES_QOs}, measured with $H \parallel c^*$ (perpendicular to the $ab$ plane); such a configuration detects the in-plane linear magnetostriction ($\alpha_{x}(H) \propto \lambda_{ab}(H)$, see Methods). Successive magnetic phase transitions occur with increasing $H$, manifest as peaks in $\alpha_{x}(H)$. At $T$ = 1.8\,K, we identify totally seven transitions up to $\mu_0H$ = 9\,T. The peaks at $\mu_0H_{\rm c1}$ = 4.9\,T and $\mu_0H_{\rm c2}$ = 7.4\,T correspond to the transitions entering and exiting the 1/3 magnetization plateau (see magnetization isotherms $M(H)$ in Fig.\,1c), respectively, in agreement with previous results \cite{Na3Ni2BiO6_plateau}. Surprisingly, two salient anomalies appear within the field range of the 1/3-plateau phase, with the maxima occurring at $\mu_{0}H_{1}^{*}$ = 5.7\,T and $\mu_{0}H_{2}^{*}$ = 6.3\,T, respectively; these features divide the plateau into three phases (phases I, II and III in Fig.\,1b), an unusual phenomenon to be discussed in more detail below. Above $\mu_0H_{\rm c2}$, the system undergoes three more transitions at $H_{\rm c3}$, $H_{\rm c4}$ and $H_{\rm c5}$, implying the presence of a remarkable cascade of intermediate phases before eventual establishment of spin saturation slightly above $\mu_0H_{\rm c5}$ = 8.2\,T. As shown in Fig.\,1d, $H_{\rm c3}$-$H_{\rm c5}$ also manifest as distinctive features in $dM/dH$; intriguingly, the dip of $dM/dH$ between $H_{\rm c3}$ and $H_{\rm c4}$ ($H_{\rm c4}$ and $H_{\rm c5}$) occurs where $M$ exhibits kinks at approximately 1/2 (2/3) of its saturation value $M_{\rm s}$, implying that the corresponding intermediate phases may also be fractional quantum states without developing a stable plateau; such behaviour is reminiscent of the multiple plateau states at higher fractions observed in triangular lattice antiferromagnets \cite{Cs2CuBr4Forture,Ba3CoSb2O9_HighField,Ba3NiNb2O9} and the Shastry-Sutherland lattice magnet SrCu$_2$(BO$_3$)$_2$ \cite{SrCuBO_Jaime,SrCuBO_Matsuda}, underlining the perplexing spin frustration in Na$_{3}$Ni$_{2}$BiO$_{6}$.

With increasing $T$, the transition fields delineating the boundaries of the 1/3 magnetization plateau show different evolution tendencies: $H_{\rm c1}$ is almost $T$ independent, whereas $H_{\rm c2}$ gradually decreases; the 1/3 magnetization plateau thus becomes narrower upon warming and is not well-defined above 6\,K (Fig.\,1c). The transitions separating the intermediate phases ($H_{\rm c3}$-$H_{\rm c5}$) shift to lower fields before merging and turning indiscernible above $\simeq$ 4\,K (Fig.\,1b). The critical fields $H_{1}^{*}$ and $H_{2}^{*}$ are largely independent of $T$ and smear out above $\simeq$ 5\,K and $\simeq$ 4\,K, respectively (Fig.\,1b).
A $H-T$ phase diagram of Na$_{3}$Ni$_{2}$BiO$_{6}$ for $T >$ 1.8\,K is presented in Fig.\,1e, which is deduced from the ac magnetostriction data, $M(H)$, torque magnetometry (Supplementary Fig.\,2) and specific heat (Supplementary Fig.\,3). The field-induced transitions $H_{\rm c1}-H_{\rm c5}$ are of first-order nature (Supplementary Sec.\,I); in Fig.\,1e, we only plot the data for the up-sweep measurements. We remark that the vertical transition lines $H_{\rm c1}$, $H_{1}^{*}$ and $H_{2}^{*}$ strongly imply that the complicated phase diagram of Na$_{3}$Ni$_{2}$BiO$_{6}$ is predominantly controlled by the zero-$T$ quantum fluctuations, instead of the entropies --- such phenomenon has been recognized as a distinctive characteristic of frustrated magnets \cite{Cs2CuBr4Forture,Na2Co2TeO6_PhaseDiagram}. This scenario is also supported by the fact that most transition lines (with the exceptions of $H_{\rm c1}$ and $H_{\rm c5}$) smear out at temperatures below $T_{\rm N}$, suggesting that the corresponding quantum phases may become unstable against thermal fluctuation before the destruction of magnetic order. Moreover, measurement of specific heat $C_{\rm p}(T)$ reveals a shoulder-like feature slightly below the sharp peak at $T_{\rm N}$ in the field range 5.1-6.8\,T (Supplementary Fig.\,3). This feature is indicative of an additional phase transition at $T^* < T_{\rm N}$ within the magnetically ordered state (the olive shaded region in Fig.\,1e); the 1/3-plateau phases evolve from the state developed below $T^*$ upon cooling (Supplementary Fig.\,5a).

\bigskip

\noindent
\textbf{The exotic 1/3-plateau phases}

\noindent
Now we further look into the multi-phase 1/3 magnetization plateau segmented by transitions $H_{1}^{*}$ and $H_{2}^{*}$ (navy shaded region in Fig.\,1e).
Whilst the 1/3 magnetization plateau has been observed in a number of triangular-lattice frustrated magnets possessing both quantum \cite{Cs2CuBr4Forture,Ba3CoSb2O9_MH,NaBaNiPO,NaBaCoPO} and classical \cite{RbFeMo2O8} spins, in such cases it is attributed to an up-up-down spin configuration \cite{StarykhReview,ChubukovPlateau,Cs2CuBr4Plateau} and the occurrence of multiple phases inside the plateau regime, as we observed here in Na$_{3}$Ni$_{2}$BiO$_{6}$, is exceptionally rare. We mention that although the transition features at $H_{1}^{*}$ and $H_{2}^{*}$ are most clearly captured by the ac magnetostriction measurement (Fig.\,1b) above 1.8\,K, thanks to its extreme sensitivity to field-induced phase transitions, corresponding signatures can also be seen in the $H$-derivatives of magnetic torque $\tau(H)$ and $M(H)$. As demonstrated in Fig.\,2a, the features in $\alpha_x$, $d(\tau/H)/dH$ and $dM/dH$ (measured on three different samples) are generally consistent in their positions, implying that the presence of phases I, II and III in the plateau regime is intrinsic to Na$_{3}$Ni$_{2}$BiO$_{6}$. Small mismatches (0.1-0.2\,T) may be due to unintentional sample misalignment with respect to $H$. These features are also reversible during up and down sweeps of $H$ (solid and dashed lines in Fig.\,2a). Hence, they are unlikely to originate from evolution of random domain structures in the 1/3 plateau phase (see Sections I and II of Supplementary Information for more details).

To clarify the nature of these exotic plateau phases, we measured $M(H)$ and $\tau(H)$ down to $T$ = 0.4\,K and 0.15\,K, respectively. Low-$T$ data reveal several additional interesting results. First, the phases I, II and III show $M/M_{\rm s} \simeq$ 0.30, (0.33-0.37) and (0.38-0.40) at 0.4\,K, respectively (Fig.\,2b); they are thus `approximate' 1/3-plateau phases" with either higher fraction or incommensurate magnetization. Second, despite that $H_{1}^{*}$ and $H_{2}^{*}$ appear to be independent of the $H$-sweep direction (Figs.\,2a and b), they are likely to be weak first-order transitions, as evidenced by the notable variation in the magnitude of $M$ for plateau phases I and II between up and down sweeps (Fig.\,2b) and the absence of signatures at $H_{1}^{*}$ and $H_{2}^{*}$ in $C_{\rm p}(H)$ (Supplementary Fig.\,3d, see Supplementary Sec.\,I). We note that the double-peak structure of the anomalies at $H_{1}^{*}$ and $H_{2}^{*}$ in $\alpha_x(H)$ (Fig.\,1b and Fig.\,2a) is possibly caused by the domain formation in the vicinity of these first-order transitions. Third, a small jump occurs between $H_{1}^{*}$ and $H_{2}^{*}$ in $\tau(H)$ for $T \leq$ 0.2\,K (red triangles in Fig.\,2c), an indication of an additional phase transition further separating phase II into IIa and IIb at the lowest $T$. The emergence of extra features upon cooling strongly challenges the conventional order-by-disorder scenario for plateau formation \cite{StarykhReview,ChubukovPlateau,Zhitomirsky,OrderByDisorder}, in which quantum fluctuation only select one of the highly-degenerate classical spin states instead of creating new phases.

The unique properties of the 1/3-plateau phases in Na$_{3}$Ni$_{2}$BiO$_{6}$ are further unveiled by contrasting the $M(T)$ and $C_{\rm p}(T)$ data at the lowest $T$. As shown in Fig.\,2d, the plateau phases exhibit distinct behaviours of $M(T)$ below $T \simeq$ 3\,K: in phase I ($\mu_0H$ = 5.3 and 5.5\,T), $M(T)$ first drops upon cooling, developing a rounded knee-shape feature before leveling off, whereas $M(T)$ in phase II (5.7-6.2\,T) and phase III (6.4-6.8\,T) show continuous decrease and nearly $T$-independent behaviour, respectively. The disparity of the slope $dM/dT$ hints at different field evolution of magnetic entropy $S(H,T)$. According to the Maxwell relation, $\partial M/\partial T = -\partial S/\partial (\mu_0H)$, the $H$-derivative of magnetic entropy is weak in phases I and III but appreciable in phase II; this inference is strengthened by the divergent magnetic Gr{\"u}neisen parameter ${\rm\Gamma}_{H}(T)=-\frac{1}{C_{\rm p}}\frac{\partial M}{\partial T}$ observed only in phase II (Fig.\,2e). ${\rm\Gamma}_{H}$ is related to the entropy as $\Delta S = \int C_{\rm p}{\rm\Gamma}_{H}d(\mu_{\rm 0}H)$ and its divergence is usually assigned to field-induced instabilities of the system, such as quantum criticality \cite{Gruneisen}. Nevertheless, we point out that although $S$ may change notably across phase II, fits of $C_{\rm p}(T)$ suggest gapped magnetic excitation spectrum in all three plateau phases (Supplementary Fig.\,6), which is consistent with nuclear magnetic resonance (NMR) \cite{Na3Ni2BiO6_NMR} results. Hence, abundance of low-$T$ magnetic entropy is unexpected for the entire (approximate) 1/3 magnetization plateau region.

\bigskip

\noindent
\textbf{Low-$T$ specific heat anomaly occurring between $H_{\rm c2}$ and full polarization}

\noindent
The specific heat data of Na$_{3}$Ni$_{2}$BiO$_{6}$ measured down to 0.1\,K for different $\mu_0H$ from 0 to 14\,T are summarized in Fig.\,3a (see Supplementary Fig.\,7 for more data). $C_{\rm p}(T)$ can be satisfyingly described by a sum of the contributions from gapped magnons ($C_{\rm m}$) and phonons ($C_{\rm ph}$) for fields below 7.5\,T (Supplementary Fig.\,6); $C_{\rm ph}$ is negligibly small below 1\,K (Supplementary Sec.\,III). Therefore, we conclude that the ground state of spin system in both the low-$H$ zigzag AFM phase and the 1/3-plateau phases (as mentioned above) are gapped, with a spin excitation gap decreasing with increasing $H$ (Supplementary Fig.\,6d). Such observations are qualitatively in agreement with a previous NMR study \cite{Na3Ni2BiO6_NMR}. The situation, however, becomes totally different when the system exits the 1/3 magnetization plateau as $H$ is further increased: whereas the NMR results suggest a gap closing point at $\mu_0H \approx$ 8.3\,T, our specific heat measurements reveal an anomalous rise of $C_{\rm p}(T)$ at the lowest $T$, which occurs in an extended field regime from 7.5 to $\sim$ 11\,T (Fig.\,3a). This is more clearly displayed in Fig.\,3b; the divergent behaviour of $C_{\rm p}(T)/T$ is completely absent at $\mu_0H$ = 7.3\,T (in plateau phase III), yet abruptly emerges once $\mu_0H > \mu_0H_{\rm c2}$ ($\approx$ 7.4\,T) and becomes the strongest at $\mu_0H \simeq$ 8.5\,T. With further increase of $H$, the divergent anomaly is suppressed above 10\,T, with $C_{\rm p}(T)/T$ at 10.5 and 10.7\,T developing a broad peak at $T_{\rm peak}$ before dropping to a very low value. Above $\mu_0H$ = 11.5\,T, the anomalous low-$T$ upturn completely vanishes (Supplementary Fig.\,7) and $C_{\rm p}(T)$ retains the form consisting of gapped magnon and phonon terms (Supplementary Fig.\,6c). We stress that the upturn persists to a field far beyond $\mu_0H_{\rm c5}$ ($\approx$ 8.2\,T, Figs.\,1e and 2b), that is, it not only exists in the intermediate phases but also apparently survives into the high-$H$ polarized state (inset of Fig.\,3b) in the phase diagram presented in Fig.\,1e. A few more intriguing characteristics of this specific heat anomaly are to be noted: (i) It only accounts for a small fraction (4-5$\%$) of the total spin entropy for a spin-1 ground state, $R\ln3$ ($R$ is the universal gas constant), between 0.1 and 1.0\,K (Fig.\,3c); (ii) the resulted excessive specific heat $C_{\rm p}(H)$ in the corresponding range of $H$ appears to be reversible upon up and down sweeps of $H$ above $H_{\rm c5}$ and becomes hysteretic below it (Supplementary Fig.\,4e); (iii) its $T$ dependence is best described by an inverse-power-law relation, $C_{\rm p}(T)/T \sim T^{-\alpha}$, $\alpha$ changes from 3.0 at 7.5\,T to 2.1 at 10.7\,T (where the fit becomes less compelling) and the most prominent anomalies are characterized by $\alpha \simeq$ 2.6-2.7, as shown in Fig.\,3d and Supplementary Fig.\,12. The peculiar field evolution and the specific values of exponent $\alpha$ exclude the nuclear or impurity Schottky contributions as the origin of this low-$T$ anomaly (Supplementary Sec.\,III, Supplementary Figs.\,12-14).

In Fig.\,4a we append the value of $C_{\rm p}/T$ to the low-$T$ magnetic phase diagram determined by torque magnetometry (Fig.\,2c and Supplementary Fig.\,2). The colour contour map unambiguously corroborates that the $C_{\rm p}$ anomaly happens immediately above $H_{\rm c2}$ at the lowest $T$. Thereby, the transition from the 1/3 plateau phase III to the first intermediate phase promptly triggers the emergence of abundant low-energy collective excitations, giving rise to the remarkable enhancement of low-$T$ specific heat. On the high-$H$ side, the red region representing high $C_{\rm p}/T$ exhibits an abnormal S-shaped boundary, at which it is cut off by an evidently gapped state with marginal $C_{\rm p}/T$ (blue region) below $\simeq$ 0.5\,K ($T_{\rm peak}$ in Fig.\,4a). More interestingly, we found that such cut-off boundary for $C_{\rm p}/T$ closely tracks the crossover line $H_{\rm p}(T)$ to the full spin polarization in this region of phase diagram; the latter is determined from $\tau(H)$ (Fig.\,4b) as the onset of high-$H$ linear behaviour (suggestive of $H$-independent magnetization, see Supplementary Sec.\,IV). The coincidence between $H_{\rm p}$ and the disappearance of $C_{\rm p}$ anomaly, as illustrated in Fig.\,4c, provides valuable information: the unusual low-$T$ upturn of $C_{\rm p}$ is closely linked to the incomplete polarization in the `nominally' polarized state above $H_{\rm c5}$,  and can be erased once all Ni spins are fully polarized by the $H$-field.

\bigskip
\noindent
\textbf{\large Discussion}

\noindent
A power-law $T$ dependence of the specific heat, $C_{\rm m}(T) \sim T^{\gamma}$ ($\gamma \lesssim$ 1), has been frequently cited as evidence for the presence of low-lying fermionic spin excitations (i.e., spinons) in frustrated quantum magnets --- which points towards the realization of gapless spin-liquid states \cite{Yamashita,Ba3CuSb2O9SL,YbCl3,Kitaev_QSL_WeiLi}. In particular, $\gamma = 1$ ($C_{\rm m} \propto T$) implies that the coherent quasiparticles form a spinon Fermi surface that is resistive against gauge field fluctuations \cite{PALeePairing}, whereas $\gamma$ = 2/3 is expected when the quasiparticle mass renormalization caused by such fluctuations is considered \cite{YbMgGaO4,Motrunich}. $T$-linear specific heat can also emerge from gapless excitations in insulating magnets when $H$ induces a QCP in transverse Ising chain magnets \cite{CoNb2O6_Tian_Liang} or a BEC-QCP in the 2D limit \cite{YbCl3,NaBaCoPO_MCE}. Alternatively, a sublinear $C_{\rm m} = T^{\beta}$ with $\beta <$ 1 may stem from bond randomness in disordered frustrated magnets \cite{H3LiIr2O6_disorder,Kimchi}. However, none of these scenarios interpret the inverse-power-law behaviour $C_{\rm p}(T)/T \sim T^{-\alpha} (\alpha > 2)$ established between $H_{\rm c2}$ and $H_{\rm p}$ in Na$_{3}$Ni$_{2}$BiO$_{6}$ (Fig.\,4a); all predictions for gapless excitations yield a much weaker divergence, $C_{\rm p}(T)/T \sim T^{-\alpha} ( 0 \leq \alpha < 1)$. 

Thereby, we propose that the enigmatically divergent $C_{\rm p}(T)/T$ indeed unveils the occurrence of a significant peak-type structure in $C_{\rm p}$ at $T <$ 100\,mK; such an anomaly most likely signals an impending transition to a masked ground-state order that, unfortunately, evades our experimental probes due to its ultra-low onset temperature. This emergent low-$T$ ordered state appears immediately once the 1/3-plateau phase III terminates and may act as the ground state for the intermediate phases between $H_{\rm c2}$ and $H_{\rm c5}$ (Fig.\,1e); note that these intermediate phases themselves are likely to be ordered with specific magnetic superstructures, as $H_{\rm c5}$ closely traces the AFM phase boundary (Fig.\,4a). More strikingly, this low-$T$ state persists beyond $H_{\rm c5}$ up to a critical field between 10 and 10.5\,T, above which its formation is cut off by the full polarization of spins that gaps out all low-lying magnons (Fig.\,4). As such, the emergent order must be essentially distinct from the AFM ordering at higher $T$ which ends at $H_{\rm c5}$ (Supplementary Sec.\,IV). A most tempting phenomenology elucidating such a field evolution considers the magnons originating from the in-plane component of the Ni spins: They can be analogously treated as interacting dilute hard-core bosons in the concerned field region and, upon cooling, these bosons potentially condense into a superfluid state at the lowest $T$. Besides the superfluidity picture, we recall that the Kitaev interaction has been proposed to play an essential role introducing the frustration in Na$_{3}$Ni$_{2}$BiO$_{6}$ \cite{Na3Ni2BiO6_plateau}; intriguingly, a pronounced peak in $C_{\rm m}$ has also been predicted to emerges in the Kitaev model at very low $T$ \cite{Kitaev_QSL_WeiLi,Kitaev_QSL_Nasu}, suggesting a crossover to a fractionalized spin-liquid ground state with aligned fluxes (still, these pure Kitaev models predict gap formation under $H$-field and are inapplicable in Na$_3$Ni$_2$BiO$_6$). Verification of the link between $C_{\rm p}$ anomaly and the Kitaev physics awaits future investigations.

The 1/3-plateau phases in Na$_{3}$Ni$_{2}$BiO$_{6}$ are also unconventional, in the sense that the plateau comprises at least three phases (maybe more at the lower $T$, Fig.\,2c) with slightly different $M$ yet distinct phase boundaries (Figs.\,1 and 2).  This marks a clear departure from familiar 1/3 magnetization plateaus in frustrated magnets, including both the classical (e.g., the plateaus occurring in triangular magnets via the order-by-disorder mechanism \cite{StarykhReview,ChubukovPlateau,OrderByDisorder,Ba3CoSb2O9_MH,Cs2CuBr4Plateau}) and the quantum ones (arising from entangled spins) \cite{SrCuBO_Jaime,SrCuBO_Matsuda}. We also note that the magnetization of plateau phases in Na$_{3}$Ni$_{2}$BiO$_{6}$ show marginal time and history dependence (Supplementary Sec.\,II), thus are distinct from the nonequilibrium magnetization steps observed in spin-chain compound Ca$_3$Co$_2$O$_6$ \cite{HardyCa3Co2O6}. Nevertheless, the gapped magnon spectrum revealed by $C_{\rm p}(T)$ (Supplementary Fig.\,6) suggests that these plateau phases are still likely to be solid states of hard-core bosons. A previous study postulated a nonclassical zero-up-zero-down-up-up spin configuration for the plateau phase, in which two out of the six spins in a magnetic unit aligned in the $ab$ plane \cite{Na3Ni2BiO6_plateau}. Starting from such model, we conjecture that the formation of plateau phases I-III have two possible origins: (i) They correspond to three distinct spin configurations, perhaps with different alignments of in-plane spin components. One likely scenario is that the two spin-flop (`zero') sites \cite{Na3Ni2BiO6_plateau} in the 6-site spin chain change their positions at the field-induced spin-reconfiguration transitions $H_{1}^{*}$ and $H_{2}^{*}$; according to our first-principles calculations (Methods and Supplementary Table 3), such site changes render the energies of spin system slightly different, thus allow the successive development of nearly-degenerate plateau phases under increasing $H$. (ii) The weak interplane exchange coupling gives rise to the segmentation of 1/3 plateau into states with different ordering vector $q$ along $c^*$, similar to the famous ``Devil's staircase" (a cascade of commensurate magnetization plateaux) in the axial next-nearest neighbour Ising (ANNNI) model \cite{ANNNI} (see Supplementary Sec.\,II for more discussions). We emphasize that the plateau phases I, II and III do not have a distinguishable upper boundary with increasing $T$ (Fig.\,1e) and their emergence completely rely on quantum fluctuations; the interplane $q$ in Na$_{3}$Ni$_{2}$BiO$_{6}$, if exist, is likely to be incommensurate.

Unexpectedly, the $H-T$ phase diagram of Na$_{3}$Ni$_{2}$BiO$_{6}$ phenomenologically emulates those of quasi-2D triangular lattice antiferromagnets which have predominant in-plane nearest-neighbor interactions \cite{NaBaCoPO,NaBaCoPO_MCE,NaBaNiPO}, in spite of the pure exchange (geometric) frustration for the former (latter). In both cases, the spin system undergoes consecutive magnetic transitions under elevated $H$, from zero-field AFM state to the 1/3 plateau phase(s), the intermediate phases with appreciable low-$T$ magnetic entropy and the gapped fully-polarized state, in the sequential order. Clear discrepancies do exist: In triangular spin lattices, the 1/3 magnetization plateau corresponds to a classical up-up-down state as mentioned above and the exotic intermediate phases are supersolid \cite{NaBaCoPO_MCE} or superfluid \cite{NaBaNiPO} states with gapless Goldstone excitations (magnons), characterized by the simultaneous breaking of U(1) gauge symmetry and an additional (e.g., translational \cite{NaBaCoPO_MCE} or spin rotational \cite{NaBaNiPO}) symmetry; in contrast, the plateau phases are of essential quantum nature in Na$_{3}$Ni$_{2}$BiO$_{6}$, and the intermediate phases beyond them cannot be standard supersolids of hard-core bosons --- for the reason that the U(1) symmetry is implicitly broken by the active Kitaev interaction --- thus the ``hidden" ground state (presumably a superfluid state, or a more exotic disordered state provoked by Kitaev physics) is not necessarily gapless. Nevertheless, the ostensible resemblance is already impressive, highlighting the astonishingly rich outcomes of Kitaev interaction in honeycomb lattices. Recent studies reveal that the Kitaev $K$ and off-diagonal $\Gamma$ interactions in an extended Kitaev model are crucial in stabilizing the 1/3-plateau phase in Na$_3$Ni$_2$BiO$_6$ (details will be presented in a forthcoming parallel paper from some of us); apart from giving rise to strong quantum fluctuations, they may as well create natural pairing terms when mapped into a bosonic representation, thus provoking the series of anomalous phenomena observed herein. Our results therefore pinpoint the great promise of bringing forth emergent spin states in honeycomb spin lattices via introducing the Kitaev coupling. Further experimental explorations down to millikelvin temperatures are required to determine the nature of 1/3-plateau and intermediate phases in Na$_{3}$Ni$_{2}$BiO$_{6}$ and we anticipate that such efforts may reveal brand-new many-body physics in frustrated spin systems.

\bigskip
\noindent
\textbf{\large Methods}

\noindent
\textbf{Sample preparation and characterizations}

\noindent
High-quality single crystals of Na$_{3}$Ni$_{2}$BiO$_{6}$ were synthesized using a Na$_{2}$CO$_{3}$-Bi$_{2}$O$_{3}$ self-flux method \cite{Na3Ni2BiO6_structure,Na3Ni2BiO6_plateau}. The typical size of the plate-shaped crystals was 4$\times$4$\times$0.3\,mm$^3$. In this work, totally five samples were studied ($\#1-\#5$, see Supplementary Sec.\,I for details). Low-temperature magnetic properties were investigated down to 0.4\,K using a Quantum Design MPMS3 system with He-3 probe. The isothermal magnetization $M(H)$ ($T \geq$ 2\,K) was measured up to 14\,T in a Quantum Design Physical Property Measurement System (PPMS-14\,T) equipped with the vibrating sample magnetometer module. Magnetic torque was measured in a DynaCool-14\,T PPMS using a piezoresistive cantilever (SCL Sensortech PRSA-L300), which converts the magnetic torque of the sample to an electrical resistance signal. The measurement configuration is shown in Supplementary Fig.\,2a. A small Na$_{3}$Ni$_{2}$BiO$_{6}$ sample ($\#4$) with dimensions of $\sim$ 0.2$\times$0.2$\times$0.05\,mm$^3$ was cut from the as-grown crystals and attached to the tip of the piezoresistive cantilever using Apiezon N grease (Supplementary Fig.\,2b). Specific heat measurements were performed in a Quantum Design PPMS (DynaCool-14\,T) utilizing a dilution refrigerator (DR) insert.

\noindent
\textbf{ac magnetostriction measurement}

\noindent
The ac magnetostriction measurement is achieved based on a composite magnetoelectric (ME) technique that is developed based on the interfacial coupling between the sample and a piezoelectric transducer \cite{MES_MnSi,MES_laminate,XiaxinDing}: the magnetostrictive response of the sample to the external ac $H$-field is sensed by the transducer via the interfacial strain, subsequently giving rise to an ac ME voltage signal across the latter due to the piezoelectric effect. Compared with conventional methods of magnetostriction measurement, this technique has the advantage that it directly measures the linear magnetostriction coefficient $\lambda$ instead of $\Delta L(H)/L_0$, thus avoids the excess noise introduced by taking the $H$ derivative of $\Delta L(H)/L_0$ and provides a particularly sensitive probe for the determination of field-induced phase transitions (we emphasize that it is so far the only experimental method which is able to capture the in-plateau transitions $H_{1}^{*}$ and $H_{2}^{*}$ at $T > $2\,K). In this work, the experiment was performed using a commercial composite ME probe (MultiField Tech.) that can be loaded into a Quantum Design PPMS. In our study, the ME composite was a Na$_{3}$Ni$_{2}$BiO$_{6}$/PMN-PT magnetoelectric laminate device (Fig.\,1a) prepared by bonding a Na$_{3}$Ni$_{2}$BiO$_{6}$ single crystal (with as-grown size) to a piezoelectric PMN-PT (0.7Pb(Mg$_{1/3}$Nb$_{2/3}$)O$_3$–0.3PbTiO$_3$) substrate with a thickness $t$ = 200\,$\mu$m using silver epoxy (Epoxy Technology, EpoTek H20E). The contacting surface of the sample was mechanically polished with fine sandpaper before mounting to improve the interfacial strain coupling of the laminate. Before experiment, the PMN-PT was pre-polarized by a 150\,V voltage supplied with a Keithley 2400 source meter. During the measurements, the superconducting magnet in the PPMS provided the dc field $H_{\rm dc}$ that controls the properties of the sample (thus noted as $H$ in this work), whereas a small ac field $H_{\rm ac} \simeq$ 2-4\,Oe superposed on $H_{\rm dc}$ was generated by a Helmholtz coil in the probe; the ac sine-wave current (with a frequency $f$ = 200-1000\,Hz) passing through the coil was supplied by a Keithley 6221 current source. The ac voltage signal $V_{ME} = V_{x}+iV_{y}$ between the top and bottom surface of PMN-PT substrate (Fig.\,1a) was collected by a Stanford Research SR830 lock-in amplifier; note that the silver epoxy between the sample and PMN-PT also served as the electrical contact. The ME voltage coefficient $\alpha_{\rm ME} = \alpha_{x}+i\alpha_{y}=(V_{x}+iV_{y})/(H_{ac}t)$ was then derived from the voltage signal. Since $V_{\rm ME}$ arises from an electric field $E_z$ perpendicular to the surface of laminate (we defined this direction as the $z$-axis of the setup, which is also the direction of both $H_{\rm dc}$ and $H_{\rm ac}$), the measured $\alpha_{\rm ME}$ was indeed $\alpha_{\rm ME}^{zz}$. According to previous studies \cite{MES_MnSi,MES_2Dmagnets,MES_laminate},
\begin{equation}
	\alpha_{\rm ME}^{zz} \approx A_{\rm ME}(\lambda^{xz}d^{xz}+\lambda^{yz}d^{yz}),
\end{equation}
where $d^{xz}$ ($d^{yz}$) is the piezoelectric coefficient of PMN-PT characterizing the electrical field along $z$ generated by deformation along $x$($y$), $\lambda^{xz}$($\lambda^{yz}$) is the transverse linear magnetostriction coefficient along $x$($y$) under $H \parallel z$ [$\lambda^{xz(yz)} = (1/L_0)\partial\Delta L_{x(y)}/\partial\mu_0H_z$], $A_{\rm ME}$ is a constant relying on sample volume and interfacial coupling. Here we took the real part of $\alpha_{\rm ME}$, i.e., $\alpha_{x}(H)$, as an indicator of the $H$-dependent in-plane linear magnetostriction $\lambda_{ab}$ for Na$_{3}$Ni$_{2}$BiO$_{6}$ \cite{MES_QOs}. Note that $\lambda_{ab}$ represents a sum of the components of $\lambda$ along two perpendicular symmetry axes in the honeycomb plane of Na$_{3}$Ni$_{2}$BiO$_{6}$; in this study we did not distinguish the in-plane directions of the Na$_{3}$Ni$_{2}$BiO$_{6}$ crystal, because magnetic domains along three equivalent axes are demonstrated to exist in this compound \cite{Na3Ni2BiO6_plateau}.

The calibration of ac magnetic field $H_{ac}$ generated by the Helmholtz coil was performed before each measurement (Supplementary Figs.\,8a and b). We replaced the sample with a small three-turn induction coil and estimated the magnitude of $H_{ac}$ by sensing the induced ac voltage signal picked up by the induction coil. $H_{ac}$ increases almost linearly as the amplitude of excitation current $I_{ac}$ raising from 20\,mA to 75\,mA (Supplementary Fig.\,8b); with the amplitude of $I_{ac}$ kept constant, we found that an increase of its frequency $f$ initially leads to a reduction of $H_{ac}$, until a saturation was reached in the high-$f$ region (Supplementary Fig.\,8a). The data presented in Fig.\,1 were collected with $f$ = 997.3\,Hz and the $\alpha_{x}(H)$ curves were obtained using the calibrated values of $H_{ac}$. We also performed testing measurements using $f$ = 211.3\,Hz and, as shown in Supplementary Fig.\,8c, no frequency dependence of the data was observed. Another factor to be considered during the ME experiment is the Joule heating from the excitation. To testify its influence, we set the system temperature at $T$ = 3\,K and applied an excitation of $I_{ac}$ = 50\,mA; the temperature drift was proved to be less than 0.1\,K, implying good thermal isolation between the sample and driven coil. Since the data collection was performed with an excitation smaller than this, we expect that our measurements were free from noticeable heating effect.

As shown in Supplementary Fig.\,8d, the identified phase transitions, most prominently $H_{\rm c2}-H_{\rm c5}$, also manifest as peak features in the imaginary part of the ME coefficient, $\alpha_{y}(H)$, though the features are weaker than those in $\alpha_{x}(H)$. It has been pointed out that the $\alpha_{y}$ signal reflects the dissipation of the sample during the piezomagnetic process \cite{MES_MnSi,MES_2Dmagnets}. The finite dissipation (with such a low excitation frequency) at the critical fields thus reflect the first-order nature of the corresponding phase transitions; moreover, the opposite signs for the signatures in $\alpha_{y}$ (negative) and $\alpha_{x}$ (positive) mimic the character of ME signals at the first-order transitions in Na$_3$Co$_2$SbO$_6$ \cite{MES_NaCoSbO}. Further analysis of $\alpha_{y}$ is beyond the scope of this work. We also perform the ME measurement with $H$ applied within the $ab$ plane of Na$_{3}$Ni$_{2}$BiO$_{6}$. Under in-plane $H$, the magnetization of Na$_{3}$Ni$_{2}$BiO$_{6}$ exhibits no fractional plateau; instead, $M(H)$ displays a step-like feature that indicates a spin-flop transition (Supplementary Fig.\,5c). Such a transition manifests itself as a predominant peak in $\alpha_{x}(H)$, as shown in Supplementary Fig.\,5d. No other signature can be resolved from the ME data with $H \parallel ab$.

\noindent
\textbf{Theoretical calculations}

\noindent
Our first-principles calculations were performed based on the density functional theory (DFT) as implemented in the Vienna ab initio simulation package (VASP) \cite{calculation_1}. The projected augmented wave (PAW) method is employed to solve the valence electron-ion interaction \cite{calculation_2}. The exchange correlation functional is described by the generalized gradient approximation (GGA) with Perdew-Burke-Ernzerhof functional \cite{calculation_3}. Based on the existing experimental results \cite{Na3Ni2BiO6_plateau}, we constructed 1$\times$3$\times$1 supercell to investigate the energy difference between different magnetic configurations. Among the four considered configurations, the 6-site spin chains in the honeycomb plane of Na$_{3}$Ni$_{2}$BiO$_{6}$ (magnetic unit of the 1/3-plateau phase proposed in Ref.\,\cite{Na3Ni2BiO6_plateau}) have two spin-flop sites (spins aligned in-plane) with varying positions, and the spins on the 4 remaining sites are aligned along the $c^{*}$ axis with three up spins and one down spins \cite{Na3Ni2BiO6_plateau}. Within the DFT+U scheme, the value of Hubbard $ U_{\rm eff} $ was chosen as 4.0\,eV for Ni 3$d$ electrons. The plane-wave-basis cutoff energy was set to 550 eV and the Brillouin zone are sampled by the 9$\times$2$\times$9 $\Gamma$-centered $k$-point mesh. The energy convergence criterion is $ 10^{-5} $\,eV, and the residual force is 0.02\,eV/\AA. Results of calculations were presented in Supplementary Table 3.

\bigskip
\noindent
\textbf{\large Data availability}

\noindent
The data underlying all the plots within the main text and the atomic coordinates used for the DFT calculations presented in Supplementary Table 3 have been deposited in the Figshare database and can be accessed at https://doi.org/10.6084/m9.figshare.32103745; 
all other data that support the findings of this study are available from the corresponding authors upon request.

\bigskip
\noindent
\textbf{\large Acknowledgements}

\noindent
We thank Zhentao Wang, Yisheng Chai, Shunli Yu, and Shang Gao for insightful discussions.
We gratefully acknowledge Jiyin Zhao, Xuguang Liu and Jun Zhou for assistance in the sample characterization performed at the Instruments Center for Physical Science, University of Science and Technology of China, and Yuyan Han for the support of magnetization measurements performed at High Magnetic Field Laboratory, HFIPS, Chinese Academy of Sciences, Hefei.
This work was financially supported by the National Key Projects for Research and Development of China with Grants Nos. 2022YFA1602600 (Z.X. and X.C.) and 2021YFA1400400 (J.W. and J.-X.L.), the National Natural Science Foundation of China with Grants Nos. 12274390 (Z.X.), 12488201 (Z.X.), 12225407 (J.W.), 12434005 (J.W.), 12574158 (Z.-Y.D.) and 12204449 (Y.Y.), the Quantum Science and Technology-National Science and Technology Major Project with Grant No. 2021ZD0302802 (Z.X. and X.C.), the Natural Science Foundation of Jiangsu Province with Grant No. BK20241250 (Y.S.) and the Basic Research Program of the Chinese Academy of Sciences Based on Major Scientific Infrastructures (Grant No. JZHKYPT-2021-08 to Z.X. and X.C.).

\bigskip
\noindent
\textbf{\large Author contributions}

\noindent
Z.X. proposed the research program. Z.X. and X.C. supervised the project. Y.S., S.B. and J.W. prepared the samples.
K.T. performed the magnetostriction, magnetization, magnetic torque and specific heat measurements with help from Z.G., Houpu Li, N.Z. and Hongyu Li.
K.T. and Z.X. analyzed the experimental data.
Y.Y., Z.-Y.D. and J.-X.L. performed the theoretical analysis and calculations.
K.T. and Z.X. wrote the paper with inputs from all authors.

\bigskip
\noindent
\textbf{\large Competing interests}

\noindent
The authors declare no competing interests.

\bigskip
\noindent
\textbf{\large Additional information}

\noindent
\textbf{Supplementary information} The online version contains
supplementary material available at https://www.nature.com/articles/s41467-026-73419-z.

\noindent
{\bf Correspondence and requests for materials} should be addressed to Jian-Xin Li, Jinsheng Wen, Ziji Xiang or Xianhui Chen.

\newpage

\begin{figure}[hbtp]
	\begin{center}
		\includegraphics[width=0.95\linewidth]{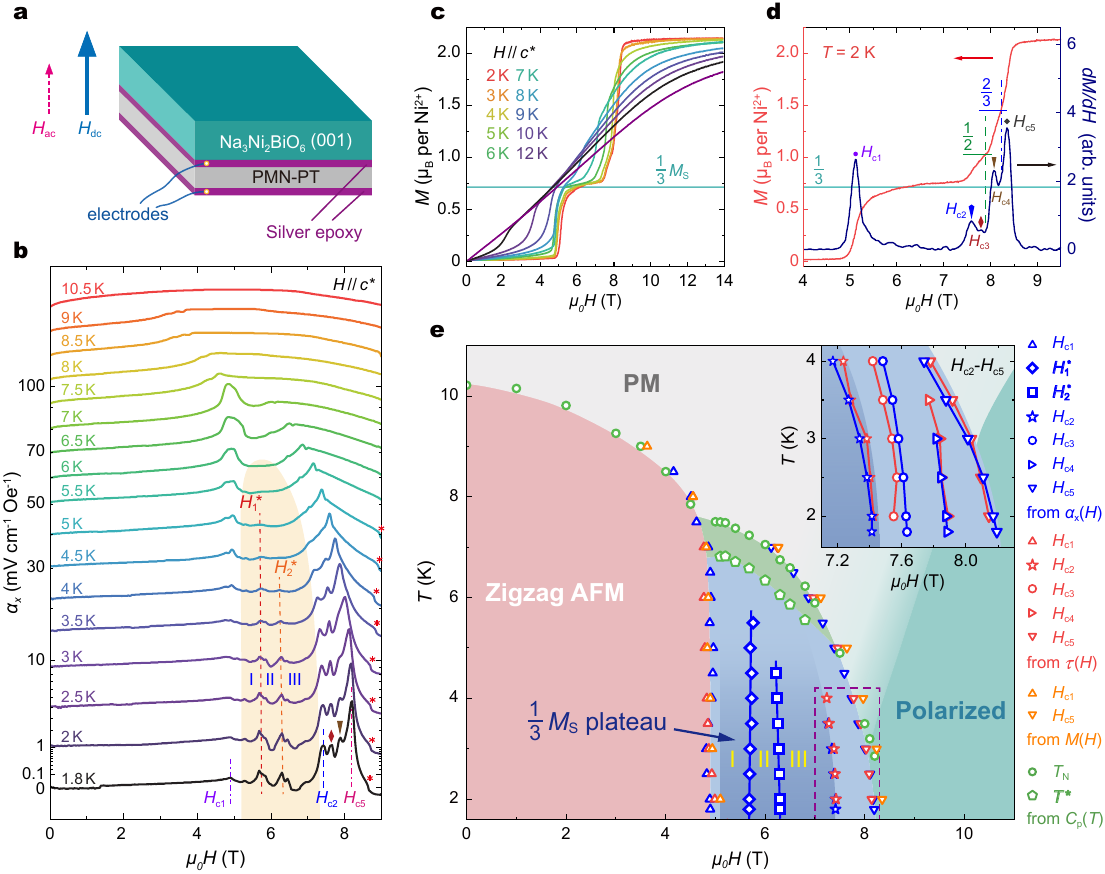}
		\caption{
			{\bf Magnetic-field-induced phase transitions and $H-T$ phase diagram for Na$_{3}$Ni$_{2}$BiO$_{6}$.}
			{\bf a,} A schematic setup of the Na$_{3}$Ni$_{2}$BiO$_{6}$/PMN-PT composite laminate used in the ac magnetostriction measurement (Methods).	
			{\bf b,} The real part of the magnetoelectric (ME) coefficient, $\alpha_{x}(H) \propto \lambda_{ab}(H)$ (see Methods), measured in Na$_{3}$Ni$_{2}$BiO$_{6}$ with $H \parallel c^*$. Orange shaded area represents the 1/3 magnetization plateau. Field induced phase transitions are marked by dashed and dash-dotted lines. Red diamonds and dark yellow down triangles indicate the unlabelled phase transitions $H_{\rm c3}$ and $H_{\rm c4}$, respectively. ``$\ast$" mark the crossover to a single-domain state. Data are plotted in the square-root coordinate and offset vertically for clarity.
			{\bf c,} Isothermal magnetization $M(H)$ ($H \parallel c^*$) for temperatures between 2 and 12\,K. Horizontal blue bar represents one-third of the saturated magnetization ($M_{\rm s}$) of Ni$^{2+}$ spins.
			{\bf d,} An expanded view of $M(H)$ and its field derivative $dM/dH$ between 4 and 9.5\,T, measured at $T$ = 2\,K. The vertical dashed lines mark out the valleys of $dM/dH$ that are potentially corresponding to higher fractional magnetization states, $ M/M_{\rm s} $ = 1/2 and 2/3.	
			{\bf e,} The $H-T$ phase diagram of Na$_{3}$Ni$_{2}$BiO$_{6}$ for $T >$ 1.8\,K under $H \parallel c^*$. Grey, magenta, navy, turquoise and olive shaded areas denote the paramagnetic phase (PM), the zigzag antiferromagnetic phase (zigzag AFM), the 1/3-plateau phase (1/3-$M_{\rm s}$ plateau, including phases I, II and III), the fully polarized state and a narrow phase regime between $T^*$ and $T_{\rm N}$, respectively. Light blue areas denote intermediate transition states surrounding the 1/3-$M_{\rm s}$ plateau. Symbols are phase transitions determined from $\alpha_{x}(H)$ (blue), magnetic torque $\tau(H)$ (red, Supplementary Fig.\,2), $M(H)$ (orange) and specific heat $C_{\rm p}(T)$ (olive, Supplementary Fig.\,3). Inset shows an expanded view of the region bounded by the dashed box.
		}
		\label{fig:figure1}
	\end{center}
\end{figure}

\begin{figure}[hbtp]
	\begin{center}
		\includegraphics[width=0.95\linewidth]{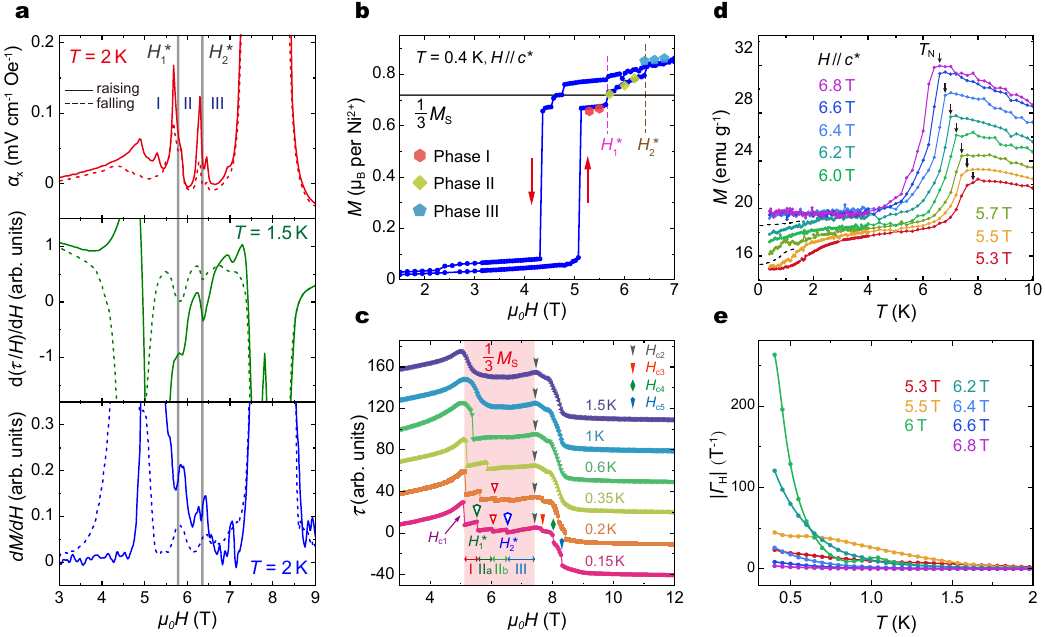}
		\caption{
			{\bf Low-temperature thermodynamic investigations of the 1/3-plateau phases.}
            {\bf a,} Comparison of the positions of anomalies in $\alpha_{x}(H)$ (upper), $d(\tau/H)/dH$ (middle) and $dM/dH$ (bottom) measured at $T$ = 1.5-2.0\,K. Solid (dashed) lines represent data collected during the field up-sweeps (down-sweeps). The vertical thin bars indicate the anomalies at transition fields $H_{1}^{*}$ and $H_{2}^{*}$.
			{\bf b,} $M(H)$ measured at 0.4\,K up to $\mu_0H$ = 7\,T. The horizontal black bar denotes the value of $M_s/3$. Red, olive and blue symbols are data points for the plateau phases I, II and III, respectively, extracted from $M(T)$ measurements under fixed $H$ (see panel {\bf d}). Arrows indicate the direction of field sweeps.
			{\bf c,} Magnetic torque curves $\tau(H)$ taken at varying $T$ between 0.15 and 1.5\,K with $H$ applied nominally along $c^*$. The field range of 1/3 magnetization plateau is highlighted by pink background. Critical fields for phase transitions beyond the plateau (see Fig.\,1e) are marked by solid symbols. Red down triangles label the step-like anomaly in $\tau(H)$ that emerges below 0.2\,K, which may represent an additional field-induced phase transition.
			{\bf d,} Zero-field-cooled (ZFC) magnetization as a function of $T$ under various $H$ in the 1/3-plateau phase regime. Dashed lines indicate the boundaries between phases I, II and III.
			{\bf e,} Magnetic Gr{\"u}neisen parameter ${\rm\Gamma}_{H}= -\frac{1}{C_{\rm p}}\frac{\partial M}{\partial T}$ plotted against $T$. Diverging behaviour upon cooling is observed only in the plateau phase II.
		}
		\label{fig:figure2}
	\end{center}
\end{figure}

\begin{figure}[hbtp]
	\begin{center}
		\includegraphics[width=0.95\linewidth]{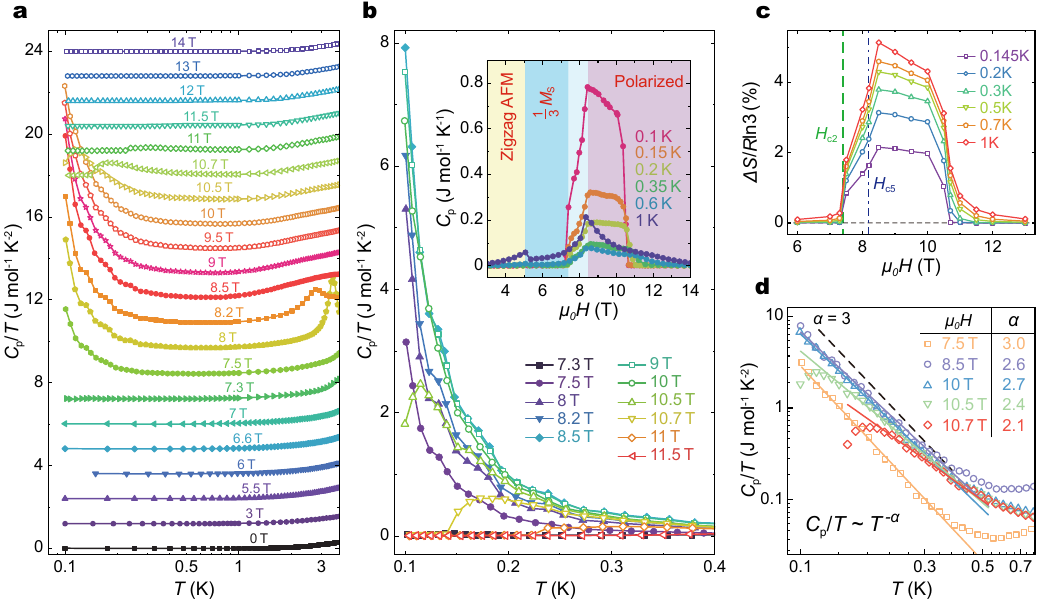}
		\caption{
			{\bf Divergent low-$T$ specific heat and its field evolution.}
			{\bf a,} Specific heat divided by temperature, $C_{\rm p}/T$, plotted against $T$ from 0.1 to 4\,K under varying $H \parallel c^*$. Data are offset vertically for clarity.
			{\bf b,} An expanded view of $C_{\rm p}(T)/T$ below 0.4\,K for 7.3\,T $\leq \mu_0H \leq$ 11.5\,T. Inset shows the $H$ dependence of $C_{\rm p}$ measured at different $T$, different background colours distinguish magnetic states: zigzag AFM (yellow), 1/3-$M_{\rm s}$ plateau (cyan), intermediate phases (light grey) and polarized state (magenta).
			{\bf c,} The $H$-dependent change of magnetic entropy, $\Delta S = S(H,T)-S(H,0.1\,{\rm K})$, calculated by integrating $C_{\rm p}/T$ with respect to $T$ from 0.1\,K to the referred temperatures. Data are normalized to $R \ln$3, where $R$ is the universal gas constant. Vertical dashed lines denote the right boundary of 1/3-plateau $H_{\rm c2}$ and highest critical field $H_{\rm c5}$ (Fig.\,1e).
			{\bf d,} Log-log plot of $C_{\rm p}/T$ versus $T$, showing the divergent features between 7.5 and 10.7\,T. Solid lines demonstrate the validity of inverse power-law fits $C_{\rm p}/T \propto T^{-\alpha}$ with $\alpha$ noted beside the curves. The black dashed line represents the expected behaviour ($\alpha = 3$) for the Schottky anomaly (Supplementary Sec.\,III).
				}
		\label{fig:figure3}
	\end{center}
\end{figure}

\begin{figure}[hbtp]
	\begin{center}
		\includegraphics[width=0.95\linewidth]{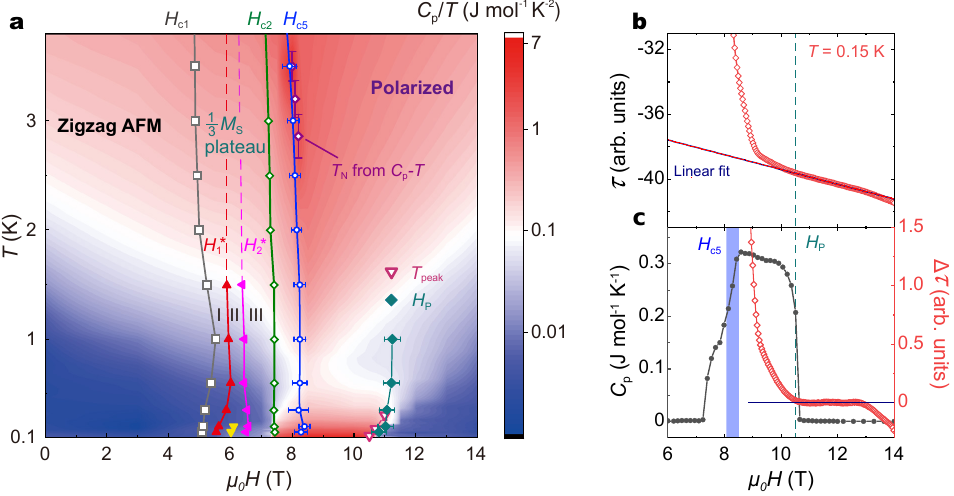}
		\caption{
			{\bf Low-temperature $H-T$ phase diagram determined from $C_{\rm p}$ and $\tau(H)$.}
			{\bf a,} Phase diagram of Na$_{3}$Ni$_{2}$BiO$_{6}$ for $H \parallel c^*$, overlaid onto the contour plots of $C_{\rm p}/T$. Open grey squares, olive diamonds and blue circles represent critical fields $H_{\rm c1}$, $H_{\rm c2}$ and $H_{\rm c5}$, respectively, determined by $\tau(H)$. Error bars of $H_{\rm c5}$ are transition widths in $d(\tau/H)/dH$ (Supplementary Fig.\,2c). Purple diamonds are $T_{\rm N}$ obtained from $C_{\rm p}(T)$ (Supplementary Fig.\,7b; errors represent the half-widths of the peaks). Within the 1/3 magnetization plateau, a cascade of transitions appear in $\tau(H)$ as denoted by triangles in red ($H_1^*$), magenta ($H_2^*$) and yellow (the small step-like anomaly develops at low $T$, Fig.\,2c). Open pink triangles indicate the broad maximum of $C_{\rm p}(T)/T$ at $T_{\rm peak}$ for fields above 10.5\,T (Supplementary Fig.\,7c). Solid cyan diamonds denote the onset field $H_{\rm p}$ for complete spin polarization, as revealed by $\tau(H)$; errors are uncertainties concerning the resolution of torque measurement. 
			{\bf b,} Linear fit for magnetic torque $ \tau(H) $ in the fully polarized state, which determines the location of $H_{\rm p}$ (Supplementary Sec.\,IV).
			{\bf c,} $H$-dependent $ \Delta\tau(H) $ obtained by subtracting linear fit shown in {\bf b}, plotted together with $C_{\rm p}(H)$. Both data were measured at 0.15\,K. Note the coincidence between $H_{\rm p}$ and the termination of specific heat anomaly.
}
		\label{fig:figure4}
	\end{center}
\end{figure}

\newpage
\renewcommand{\thefigure}{\textbf{\arabic{figure}}}
\renewcommand{\thetable}{\textbf{\arabic{table}}}
\setcounter{figure}{0}
\renewcommand{\figurename}{\textbf{Supplementary Fig.\,$\!\!$}}
\renewcommand{\tablename}{\textbf{Supplementary Table\,$\!\!$}}

\clearpage

\newpage

\newpage
\clearpage

\newpage

\end{document}